\documentclass{amsart}
\usepackage{amsfonts}

\usepackage{graphicx}

\input{tcilatex}
\input tcilatex

\begin{document}
\title{A Bose-Einstein Approach to the Random Partitioning of an Integer}
\author{Thierry E. Huillet}
\address{Laboratoire de Physique Th\'{e}orique et Mod\'{e}lisation\\
CNRS-UMR 8089 et Universit\'{e} de Cergy-Pontoise\\
2 Avenue Adolphe Chauvin, F-95302, Cergy-Pontoise, France\\
E-mail: Thierry.Huillet@u-cergy.fr}
\maketitle

\begin{abstract}
Consider $N$ equally-spaced points on a circle of circumference $N$. Choose
at random $n$ points out of $N$ on this circle and append clockwise an arc
of integral length $k$ to each such point. The resulting random set is made
of a random number of connected components. Questions such as the evaluation
of the probability of random covering and parking configurations, number and
length of the gaps are addressed. They are the discrete versions of similar
problems raised in the continuum. For each value of $k$, asymptotic results
are presented when $n,N$ both go to $\infty $ according to two different
regimes. This model may equivalently be viewed as a random partitioning
problem of $N$ items into $n$ recipients. A grand-canonical balls in boxes
approach is also supplied, giving some insight into the multiplicities of
the box filling amounts or spacings. The latter model is a $k-$nearest
neighbor random graph with $N$ vertices and $kn$ edges. We shall also
briefly consider the covering problem in the context of a random graph model
with $N$ vertices and $n$ (out-degree $1$) edges whose endpoints are no more
bound to be neighbors.\newline

\textbf{Running title:} Bose-Einstein and Integer Partitioning\newline

\textbf{Keywords:} Random integer partition, random allocation, discrete
covering of the circle, discrete spacings, balls in boxes, Bose-Einstein, $k-
$nearest neighbor random graph.\newline
\end{abstract}

\section{Introduction}

Many authors considered the problems related to the coverage of the unit
circle by arcs of equal sizes randomly placed on the circle, among which 
\cite{With}, \cite{Stevens}, \cite{Feller}, \cite{Flatto}, \cite{Siegel}, 
\cite{Holst2}, \cite{Holst3}, \cite{Hui}. In this Note, motivated by a
Remark in the paper (\cite{Can}, p.$18$) on random graphs, we shall be
concerned by a discrete version to the above problem, following \cite{Holst4}
and \cite{Iv}: Consider $N$ equally spaced points (vertices) on the circle
of circumference $N$ so with arc length $1$ between consecutive points.
Sample at random $n$ out of these $N$ points and consider the discrete
random spacings between consecutive sampled points, turning clockwise on the
circle. Let $k$ be an integer and append clockwise an arc of length $k$ to
each sampled points, forming a random set of arcs on the circle. What is the
probability that the circle is covered? If the circle is not covered, how
many gaps do we have in the random set of arcs? What is the probability that
no arc overlap (the discrete hard rods model), what is the probability that
no arc overlap and that the gaps lengths are smaller than $k$ itself (the
discrete version of R\'{e}nyi's parking model). All these questions require
some understanding of both the smallest and largest spacings in the sample.
This model can equivalently be formulated in terms of the random
partitioning of $N$ items into $n$ recipients. Here also, the distributions
of the smallest and largest shares attached to each of the recipients are of
fundamental interest. We will focus on the thermodynamical limit regime: $%
n,N\rightarrow \infty $ while $n/N\rightarrow \rho $ and also, sometimes, in
a regime where $n,N\rightarrow \infty $, while $n\left( 1-\frac{n}{N}\right)
^{k}\rightarrow \alpha $, $0<\alpha <\infty $. In the first regime, the
occurrence of say covering and parking configurations are exponentially rare
in the whole admissible density range of $\rho $, whereas in the second one
they are macroscopically frequent. At the heart of these models is the
Bose-Einstein distribution for discrete spacings. Finally, a Bosonic grand
canonical approach to the above model will be considered where $N$ balls are
assigned at random to $N$ boxes. For this urn model, we will study the
number of empty boxes and the number of boxes with $i$ balls, giving some
insight into the spacings multiplicities, both in the canonical and the
grand-canonical ensembles.\newline

The model just developed is a $k-$nearest neighbors random graph with $N$
vertices and $kn$ edges. In the last Section, we consider a random graph
with $N$ vertices and $n$ (out-degree $1$) edges whose endpoints are no more
necessarily neighbors, being now chosen at random on the whole set of
vertices. In this model of a different kind, each of the $n$ sampled points
is allowed to create a link far away with any of the $N$ vertices, not
necessarily with neighbors. We estimate the covering probability for this
random graph model in the spirit of Erd\H{o}s-R\'{e}nyi (see \cite{Bo}). We
show that, in sharp contrast to the $k-$nearest neighbor graph, there exists
a critical density $\rho _{c}=1-e^{-1}$ above which covering occurs with
probability one. The take-home message is to what extent when connections
are not restricted to neighbors, the chance of connectedness is increased.

\section{Random partition of an integer and discrete spacings}

Consider a circle of circumference $N,$ with $N$ integer. Consider $N$
equally spaced points on the circle so with arc length $1$ between
consecutive points. We shall call this discrete set of points the $N$%
-circle. Draw at random $n\in \left\{ 2,..,N-1\right\} $ points without
replacement at the integer sites of this circle (thus, with $M_{1},..,M_{n}$
independent and identically distributed, say iid,\emph{\ }and uniform on $%
\left\{ 1,..,N\right\} $). Pick at random one the points $M_{1},..,M_{n}$
and call it $M_{1:n}$. Next, consider the ordered set of integer points $%
\left( M_{m:n}\text{, }m=1,..,n\right) $, turning clockwise on the circle,
starting from $M_{1:n}$. Let $N_{m,n}=M_{m+1:n}-M_{m:n}$, $m=1,..,n-1$, be
the consecutive discrete spacings, with $N_{n,n}=M_{1:n}-M_{n:n}$, modulo $N$%
, closing the loop. Under our hypothesis, $N_{m,n}\overset{d}{=}N_{n}$, $%
m=1,..,n$, independent of $m$, the distribution of which is $\overline{F}%
_{N_{n}}\left( k\right) :=\mathbf{P}\left( N_{n}>k\right) =1-F_{N_{n}}\left(
k\right) =\binom{N-k-1}{n-1}/\binom{N-1}{n-1}$, with $\mathbf{E}N_{n}=N/n.$%
\newline

It is indeed a result of considerable age (see e.g. \cite{Holst4}) that
identically distributed (id)\emph{\ }discrete\emph{\ }spacings $\mathbf{N}%
_{n}:=\left( N_{m,n};m=1,..,n\right) $, with $\left| \mathbf{N}_{n}\right|
:=\sum_{m}N_{m,n}=N$ can be generated as the conditioning 
\begin{equation}
\mathbf{N}_{n}=\mathbf{G}_{n}\mid \left\{ \left| \mathbf{G}_{n}\right|
=N\right\} ,  \label{a1}
\end{equation}
where $\left| \mathbf{G}_{n}\right| :=\sum_{m=1}^{n}G_{m}$ is the sum of $n$
iid\textit{\ }geometric$\left( \alpha \right) $ random variables $\geq 1$
(with $\mathbf{P}\left( G_{1}\geq k\right) =\alpha ^{k-1},$ $k\geq 1$, $%
\alpha \in \left( 0,1\right) $) and so $N_{n}$ has the claimed
P\`{o}lya-Eggenberger \emph{PE}$\left( 1,n-1\right) $ distribution: $\mathbf{%
P}\left( N_{n}=k\right) =\binom{N-k-1}{n-2}/\binom{N-1}{n-1}$, $k=1,..,N-n+1.
$

Note that as $n,N\rightarrow \infty $, while $n/N=\rho <1$ is fixed, using
Stirling formula, we get the convergence in distribution 
\begin{equation}
N_{n}\overset{d}{\rightarrow }G,  \label{a2}
\end{equation}
where $G\geq 1$ is a discrete random variable (rv) with geometric$\left(
1-\rho \right) $ distribution: $\mathbf{P}\left( G\geq m\right) =\left(
1-\rho \right) ^{m-1},$ $m\geq 1.$ The limiting expected value of $N_{n}$ is 
$1/\rho .$

With $\mathbf{k}:=\left( k_{m};m=1,..,n\right) $, the joint law of $\mathbf{N%
}_{n}$ is 
\begin{equation}
\mathbf{P}\left( \mathbf{N}_{n}=\mathbf{k}\right) =\frac{1}{\binom{N-1}{n-1}}%
1\left( \left| \mathbf{k}\right| =N\right) ,  \label{a3}
\end{equation}
which is the exchangeable uniform distribution on the restricted discrete $%
N- $simplex $\left| \mathbf{k}\right| :=\sum_{m=1}^{n}k_{m}=N$, $k_{m}\geq 1$%
, also known as the Bose-Einstein distribution$.$ This distribution occurs
in the following P\`{o}lya-Eggenberger urn model context (see \cite{John}):
An urn contains $n$ balls all of different colors.. A ball is drawn at
random and replaced together while adding another ball of the same color..
Repeating this $N-n$ times, $\mathbf{N}_{n}$ is the number of balls of the
different colors in the urn. See \cite{Holst4}.

From the random model just defined, we get, 
\begin{equation}
N=\sum_{m=1}^{n}N_{m,n}  \label{a4}
\end{equation}
which corresponds to a random partition of $N$ into $n$ id parts or
components $\geq 1.$\newline

It also models the following random allocation problem (see \cite{Kol}): $N$%
\ items are to be shared at random between $n$\ recipients. $N_{m,n}$ is the
amount of the $N$\ items allocated to recipient $m$. Although all shares are
id, there is a great variability in the recipients parts as it will become
clear from the detailed study of the smallest and largest shares in the
sample.\newline

This model is connected to the continuous spacings between $n$ randomly
placed points on the unit circle in the following way: As $N\rightarrow
\infty $, $\mathbf{N}_{n}/N\overset{d}{\rightarrow }\mathbf{S}_{n}$ where $%
\mathbf{S}_{n}:=\left( S_{1,n},..,S_{n,n}\right) $ has Dirichlet uniform
density function on the continuous unit $n-$simplex \cite{Pyke} 
\begin{equation}
f_{S_{1},..,S_{n}}\left( s_{1},..,s_{n}\right) =\left( n-1\right) !\cdot
\delta _{\left( \sum_{m=1}^{n}s_{m}-1\right) }.  \label{eq0}
\end{equation}

Let $P_{n}\left( 1\right) :=\sum_{m=1}^{n}1\left( N_{m,n}>1\right) $ be the
amount of sampled points whose distance to their clockwise neighbors is more
than one unit. There are $n-P_{n}\left( 1\right) $ sampled points which are
neighbors, therefore 
\begin{eqnarray*}
N &=&1\cdot \left( n-P_{n}\left( 1\right) \right)
+\sum_{m=1}^{n}N_{m,n}1\left( N_{m,n}>1\right) \\
&=&n+\sum_{m=1}^{n}\left( N_{m,n}-1\right) _{+}
\end{eqnarray*}
where $i_{+}=\max \left( i,0\right) .$ Appending an arc of length $1$
clockwise to the $n$ sampled points and considering the induced covered set
from $\left\{ 1,..,N\right\} $, $\overline{\mathcal{L}}_{n}\left( 1\right)
:=\sum_{m=1}^{n}\left( N_{m,n}-1\right) _{+}$ represents the length of the
gaps (the size of the uncovered set). So, from the model $\overline{\mathcal{%
L}}_{n}\left( 1\right) =N-n$ is a constant and 
\begin{equation*}
N-n=\sum_{m=1}^{n}\left( N_{m,n}-1\right) _{+} 
\end{equation*}
corresponds to a random partition of $N-n$ into $n$ id parts or components $%
\geq 0$. Stated differently, the length of the covered set $\mathcal{L}%
_{n}\left( 1\right) =N-\overline{\mathcal{L}}_{n}\left( 1\right) $ is
constant equal to $n$, which is obvious$.$\emph{\newline
}

Of considerable interest is the sequence $\left( N_{m:n};m=1,..,n\right) $
obtained while ordering the components sizes $\left( N_{m,n};m=1,..,n\right) 
$, with $N_{1:n}\leq ..\leq N_{n:n}$.

By the exclusion-inclusion principle, the cumulative distribution function $%
F_{N_{m:n}}\left( k\right) =\mathbf{P}\left( N_{m:n}\leq k\right) $ is
easily seen to be 
\begin{equation}
F_{N_{m:n}}\left( k\right) =\frac{1}{\binom{N-1}{n-1}}\sum_{q=m}^{n}\binom{n%
}{q}\sum_{p=n-q}^{n}\left( -1\right) ^{p+q-n}\binom{q}{n-p}\binom{N-pk-1}{n-1%
}  \label{a5}
\end{equation}
which has been known for a while in the context of spacings in the continuum
(see \cite{With}).

In particular, 
\begin{equation}
F_{N_{n:n}}\left( k\right) :=\mathbf{P}\left( N_{n:n}\leq k\right) =\frac{1}{%
\binom{N-1}{n-1}}\sum_{p=0}^{n}\left( -1\right) ^{p}\binom{n}{p}\binom{N-pk-1%
}{n-1}  \label{a6}
\end{equation}
and 
\begin{equation}
\overline{F}_{N_{1:n}}\left( k\right) :=\mathbf{P}\left( N_{1:n}>k\right) =%
\binom{N-nk-1}{n-1}/\binom{N-1}{n-1}  \label{a7}
\end{equation}
are the largest and smallest component sizes distributions in this case.

In the formula giving $F_{N_{n:n}}\left( k\right) ,$ with $\left[ x\right] $
standing for the integral part of $x$, the sum should as well stop at $%
n\wedge \left[ \frac{N-n}{k}\right] $, observing $\binom{i}{j}=0$ if $i<j$.%
\newline

Clearly, if $k=1$, $\mathbf{P}\left( N_{n:n}=1\right) =0$ ($=1$) whatever $%
n<N$ (if $n=N$). If $k=2$ and $N>2n$, $\mathbf{P}\left( N_{n:n}\leq 2\right)
=\mathbf{P}\left( N_{n:n}=2\right) =0.$ If $N=2n$, $\mathbf{P}\left(
N_{n:n}=2\right) =1/\binom{2n-1}{n-1}$ is the probability of a regular
configuration with all sampled points equally-spaced by two arc length
units. If $n<N<2n,$ $\mathbf{P}\left( N_{n:n}=2\right) $ is the probability
of a configuration with $2n-N$ neighbor points distant of one arc length
unit and $N-n$ points distant of two units.

As $N,k\rightarrow \infty $ while $k/N\rightarrow s$%
\begin{equation*}
\binom{N-pk-1}{n-1}/\binom{N-1}{n-1}\rightarrow \left( 1-ps\right)
_{+}^{n-1} 
\end{equation*}

With $0<a<b\leq N,$ the joint law of $\left( N_{1:n},\text{ }N_{n:n}\right) $
is given by 
\begin{equation}
\mathbf{P}\left( N_{1:n}>a,N_{n:n}\leq b\right) =\sum_{m=0}^{n}\frac{\left(
-1\right) ^{m}}{\binom{N-1}{n-1}}\binom{n}{m}\binom{N-\left( na+m\left(
b-a\right) \right) -1}{n-1}.  \label{a8}
\end{equation}
In the random partitioning of $N$ image, it gives the probability that the
shares of all $n$ recipients all range between $a$ and $b.$ Putting $\left(
a=k,b=N\right) $ and $\left( a=0,b=k\right) $ gives $F_{N_{n:n}}\left(
k\right) $ and $\overline{F}_{N_{1:n}}\left( k\right) .$ This formula was
first obtained by \cite{Darling} in the continuum. Putting next $a=k$, $b=2k$%
, we get 
\begin{equation}
\mathbf{P}\left( N_{1:n}>k,\text{ }N_{n:n}\leq 2k\right) =\frac{1}{\binom{N-1%
}{n-1}}\sum_{m=0}^{n}\left( -1\right) ^{m}\binom{n}{m}\binom{N-\left(
n+m\right) k-1}{n-1}.  \label{a9}
\end{equation}
When $k=1,$ we have $\mathbf{P}\left( N_{1:n}>1,\text{ }N_{n:n}\leq 2\right)
=\binom{N-1}{n-1}^{-1}1_{N=2n}$. If $N=2n$, $\binom{2n-1}{n-1}^{-1}$ is the
probability of the configuration where the $n$ sampled points are exactly
equally-spaced, each by two arc length units.

\emph{\ }As $n,N\rightarrow \infty $ while $n/N\rightarrow \rho <1,$ with $E$
a rv with rate $1$ exponential distribution 
\begin{equation}
-\log \left( 1-\rho \right) n\left( N_{1:n}-1\right) \overset{d}{\rightarrow 
}E;\text{ }\frac{\rho }{\log n}N_{n:n}\overset{a.s.}{\rightarrow }1,
\label{a10}
\end{equation}
suggesting that the smaller (larger) integer component in the partition of $%
N $ is of order $n^{-1}$ (respectively $\log n$) in the considered
asymptotic regime. More precisely, using the joint law of $\left( N_{1:n},%
\text{ }N_{n:n}\right) $ 
\begin{equation}
\left( -\log \left( 1-\rho \right) n\left( N_{1:n}-1\right) ,N_{n:n}-\frac{%
\log n}{\rho }\right) \overset{d}{\rightarrow }\left( E,G\right) ,
\label{a11}
\end{equation}
where $\left( E,G\right) $ are independent rvs on $\Bbb{R}_{+}\times \Bbb{R}$
with distributions $\mathbf{P}\left( E>t\right) =e^{-t}$ and $\mathbf{P}%
\left( G\leq t\right) =e^{-e^{-t}}$ with $\mathbf{E}\left( G\right) =\gamma ,
$ the Euler constant (exponential and Gumbel).

Although in the random partitioning of $N$, all parts attributed to each
recipient are id, there is a great variability in the shares as the smallest
one is of order $1$ and the largest one of order $\log n.$

\section{$N-$circle covering problems}

Let $\mathcal{S}_{n}:=\left\{ M_{1},..,M_{n}\right\} $ be the discrete set
of points drawn at random on the $N-$circle with circumference $N$. Fix $%
k\in \left\{ 1,..,N\right\} $. Consider the coarse-grained discrete random
set of intervals 
\begin{equation}
\mathcal{S}_{n}\left( k\right) :=\left\{ M_{1}+l,..,M_{n}+l,1\leq l\leq
k\right\}  \label{eq1}
\end{equation}
appending clockwise an arc of integral length $k\geq 1$ to each
starting-point atom of $\mathcal{S}_{n}$.\newline

\textbf{The number of gaps and the length of the covered set.} Let $%
P_{n}\left( k\right) $ be the number of gaps of $\mathcal{S}_{n}\left(
k\right) $ (which is also the number of connected components), so with $%
P_{n}\left( k\right) =0$ as soon as the $N-$circle is covered by $\mathcal{S}%
_{n}\left( k\right) $.

Let also $\mathcal{L}_{n}\left( k\right) $ be the total integral length of $%
\mathcal{S}_{n}\left( k\right) $. As there are $n-P_{n}\left( k\right) $
spacings covered by $k$ and $P_{n}\left( k\right) $ gaps each contributing
of $k$ to the covered length, it can be expressed as a contribution of two
terms ($i\wedge j=\min \left( i,j\right) $), 
\begin{equation}
\mathcal{L}_{n}\left( k\right) =\sum_{m=1}^{n-P_{n}\left( k\right)
}N_{m:n}+kP_{n}\left( k\right) =\sum_{m=1}^{n}\left( N_{m,n}\wedge k\right) .
\label{a12}
\end{equation}
Note also that the vacancy, which is the length of the $N$-circle not
covered by any arc is 
\begin{equation}
\overline{\mathcal{L}}_{n}\left( k\right) :=N-\mathcal{L}_{n}\left( k\right)
=\sum_{p=1}^{P_{n}\left( k\right) }\left( M_{n-p+1:n}-k\right)
=\sum_{m=1}^{n}\left( N_{m,n}-k\right) _{+},  \label{a13}
\end{equation}
summing the gaps' lengths over the gaps (with $N_{n:n}-k$ the largest gaps
size and $N_{n-P_{n}\left( k\right) +1:n}-k$ the smallest gaps' size). We
recover the result $\left( i\right) $ originally due to \cite{Stevens} and
its asymptotic consequences. The following statements are mainly due to
Holst, see \cite{Holst4}. It holds that\newline

$\left( i\right) $ The distribution of $P_{n}\left( k\right) $ is 
\begin{equation}
\mathbf{P}\left( P_{n}\left( k\right) =p\right) =\frac{\binom{n}{p}}{\binom{%
N-1}{n-1}}\sum_{m=p}^{n}\left( -1\right) ^{m-p}\binom{n-p}{m-p}\binom{N-mk-1%
}{n-1}.  \label{eq2}
\end{equation}

$\left( ii\right) $\ As $n,N\rightarrow \infty $, while $n\left( 1-\frac{n}{N%
}\right) ^{k}\rightarrow \alpha $, $0<$\ $\alpha <\infty $%
\begin{equation}
P_{n}\left( k\right) \rightarrow \text{Poi}\left( \alpha \right) ,
\label{eq2a}
\end{equation}
where Poi$\left( \alpha \right) $\ is a random variable with Poisson
distribution of parameter $\alpha $.

$\left( iii\right) $\ 

$a.$ Number of gaps. As $n,N\rightarrow \infty $\ $0$\ while $n/N\rightarrow
\rho ,$\ with $0<\rho <1$, 
\begin{equation}
\frac{1}{\sqrt{n}}\left( P_{n}\left( k\right) -n\left( n/N\right)
^{k}\right) \overset{d}{\underset{N\rightarrow \infty }{\rightarrow }}%
\mathcal{N}\left( 0,\sigma ^{2}=\rho ^{k}\left( 1-\rho ^{k}\right) \right) .
\label{eq2b}
\end{equation}
where $\mathcal{N}\left( m,\sigma ^{2}\right) $\ stands for the normal law
with mean $m$\ and variance $\sigma ^{2}$.

$b.$ Gap length: 
\begin{equation}
\frac{1}{\sqrt{n}}\left( \overline{\mathcal{L}}_{n}\left( k\right) -N\left(
1-n/N\right) ^{k}\right) \overset{d}{\underset{N\rightarrow \infty }{%
\rightarrow }}\mathcal{N}\left( 0,\sigma ^{2}\right)  \label{eq2c}
\end{equation}
where $\sigma ^{2}=\left( 1+\overline{\rho }-\overline{\rho }^{k}\right) 
\overline{\rho }^{k}-\left( \overline{\rho }+k\rho \right) ^{2}\overline{%
\rho }^{2k-1}$, $\overline{\rho }=1-\rho .$\newline

The proofs of $\left( ii\right) $ and $\left( iiib\right) $ are in \cite
{Holst4}. The one of $\left( iiia\right) $\ follows from similar Central
Limit Theorem arguments developed there.\emph{\ }In the first case $\left(
ii\right) $, $n\sim N\left( 1-\left( \frac{\alpha }{N}\right) ^{1/k}\right) $
and so $n$ is very close to $N:$ because of that, there are finitely many
gaps in the limit and the covering probability is $e^{-\alpha },$ so
macroscopic. Whereas in the second case $\left( iii\right) $, $n\sim \rho N$
is quite small: the number of gaps is of order $n\rho ^{k}$ and the covering
probability is expected to be exponentially small$.$\emph{\ }Note from $%
\left( iiib\right) $ that the variance of the limiting normal law is $0$
when $k=1,$ in accordance with the fact that $\overline{\mathcal{L}}%
_{n}\left( 1\right) =N-n$ remains constant.\emph{\ }$\diamond $\emph{\newline
}

\textbf{The number of arcs needed to cover the }$N$\textbf{-circle.} In (\ref
{eq2}), $\mathbf{P}\left( P_{n}\left( k\right) =0\right) $ is the cover
probability and $\mathbf{P}\left( P_{n}\left( k\right) =n\right) $ the
probability that no overlap of arcs or rods takes place (the hard rods
model). We have $\mathbf{P}\left( P_{n}\left( k\right) =0\right) =\mathbf{P}%
\left( N_{n:n}\leq k\right) .$

The cover probability $\mathbf{P}\left( P_{n}\left( k\right) =0\right) $ is
also the probability that the number of arcs of length $k$ (the sample
size), say $N\left( k\right) $, required to cover the $N-$circle is less or
equal than $n$. We have $N\left( k\right) =\inf \left( n:N_{n:n}\leq
k\right) $. In other words, $\mathbf{P}\left( N\left( k\right) >n\right) =%
\mathbf{P}\left( N_{n:n}>k\right) $ and so $\mathbf{E}N\left( k\right)
=\sum_{n=1}^{N}\mathbf{P}\left( N_{n:n}>k\right) $, with 
\begin{equation*}
\mathbf{P}\left( N_{n:n}>k\right) =\frac{1}{\binom{N-1}{n-1}}%
\sum_{m=1}^{n}\left( -1\right) ^{m-1}\binom{n}{m}\binom{N-mk-1}{n-1}. 
\end{equation*}
We wish to estimate $\mathbf{E}N\left( k\right) $ as $N$ grows large.

When $n\left( 1-\frac{n}{N}\right) ^{k}\rightarrow \alpha $, so when $n\sim
N\left( 1-\left( \frac{\alpha }{N}\right) ^{1/k}\right) $, we have $\mathbf{P%
}\left( P_{n}\left( k\right) =0\right) =\mathbf{P}\left( N\left( k\right)
\leq n\right) \rightarrow e^{-\alpha }.$ Therefore, as $N\rightarrow \infty $
\begin{equation}
N^{1/k}\left( 1-\frac{N\left( k\right) }{N}\right) \overset{d}{\rightarrow }%
E_{k},  \label{eq3}
\end{equation}
where $E_{k}$ has a Weibull$\left( k\right) $ distribution with $\mathbf{P}%
\left( E_{k}>x\right) =e^{-x^{k}}$ and $\mathbf{E}\left( E_{k}\right)
=\Gamma \left( 1+k^{-1}\right) .$ Thus 
\begin{equation}
\mathbf{E}N\left( k\right) \sim _{N\rightarrow \infty }N\left( 1-\frac{%
\Gamma \left( 1+k^{-1}\right) }{N^{1/k}}+o\left( N^{-1/k}\right) \right)
\label{eq4}
\end{equation}
is the estimated expected number of length-$k$ arcs required to cover the $%
N- $circle.

\section{Large deviation rate functions in the thermodynamical limit: Hard
rods, covering and parking configurations}

$k-$Hard rods configurations are those for which $N_{1:n}>k\geq 1$ (the
smallest part in the decomposition of $N$ exceeds the arc-length $k:$
appending an arc of length $k$ to all sampled points does not result in
overlapping of the added arcs)$.$ $k-$Covering configurations with $k>1$ are
those for which $N_{n:n}\leq k$ (the largest part in the decomposition of $N$
is smaller than arc-length $k:$ appending an arc of length $k$ to each
sampled points results in the covering of all points of the $N$-circle, a
connectedness property)$.$ $k-$Parking configurations are those for which
both $\left( N_{1:n}>k\text{ and }N_{n:n}\leq 2k\right) $ (the smallest part
in the decomposition of $N$ exceeds the arc-length $k$ and the largest part
in the decomposition of $N$ is smaller than twice the arc-length $k:$
appending an arc of length $k$ to all sampled points results in a hard rods
configuration where sampled points are separated by gaps of length at least $%
k$ but with the extra excess gaps being smaller than $k,$ so with no way to
add a new rod (or car) with size $k$ without provoking an overlap)$.$ All
these configurations are exponentially rare in the thermodynamic limit $%
n,N\rightarrow \infty $ while $n/N\rightarrow \rho \in \left( 0,1\right) .$
We make precise this statement by computing the large deviation rate
functions in each case, extending to the discrete formulation similar
results obtained in the continuum, see \cite{Hui1}.

\subsection{Hard rods}

$k-$hard rods configurations are those for which $\left( N_{1:n}>k>1\right) $
[In the partitioning approach of the fortune $N$ amongst $n$ recipients,
this event is realized if the share of the poorest is bounded below by $k$,
a rare event]. When the number of sampled points $n$ is a fraction of $N$
(the case with a density $n=\rho N$), there are too few sampled points for a
non-overlapping configuration to occur with a reasonably large probability.
Rather, one expects that the probability of non-overlapping (hard-rods)
configurations tends to zero exponentially fast. To see this, we need to
evaluate the large $n$ expansion of $\mathbf{P}\left( N_{1:n}>k\right) $.
Note that the event $N_{1:n}>k$ is an event with positive probability if and
only if $N\geq n\left( k+1\right) $ so, in the sequel, we shall assume that $%
\rho <1/\left( k+1\right) $, $k\geq 1$. We have 
\begin{equation*}
\mathbf{P}\left( N_{1:n}>k\right) =\frac{Z_{n,N}}{\sum_{k_{1},..,k_{n}\geq
1}\prod_{m=1}^{n}1_{\sum k_{m}=N}}=\frac{Z_{n,N}}{\binom{N-1}{n-1}}\sim
C\left( \frac{n}{N}\right) ^{n}\left( 1-\frac{n}{N}\right) ^{N-n}Z_{n,N} 
\end{equation*}
where $Z_{n,N}=\sum_{k_{1},..,k_{n}\geq 1}\prod_{m=1}^{n}1_{k_{m}>k}1_{\sum
k_{m}=N}.$ In the limit $n,N\rightarrow \infty $ with $n/N\rightarrow \rho ,$
\begin{equation}
-\frac{1}{n}\log \mathbf{P}\left( N_{1:n}>k\right) \rightarrow -\frac{1}{%
\rho }\left( \rho \log \rho +\left( 1-\rho \right) \log \left( 1-\rho
\right) \right) +\lim_{n\rightarrow \infty }-\frac{1}{n}\log Z_{n,N}.
\label{b1}
\end{equation}

In the limit $n,N\rightarrow \infty $ with fixed $n/N$ limit, the quantity $%
\mathbf{P}\left( N_{1:n}>k\right) $ is easier to evaluate in an isobaric
ensemble where the pressure $p$ is held fixed instead of $\sum k_{m}.$
Therefore, relaxing the constraint $\sum k_{m}=N$, we shall work instead
with the modified random variables $\widetilde{N}_{m,n}$, with exponentially
tilted law 
\begin{equation*}
\mathbf{P}\left( \widetilde{N}_{m,n}=k_{m}>k,m=1,..,n\right) =\frac{%
\prod_{m=1}^{n}1_{k_{m}>k}e^{-pk_{m}}}{Z_{n,p}}. 
\end{equation*}
Here 
\begin{equation*}
Z_{n,p}=\sum_{k_{1},..,k_{n}\geq
1}\prod_{m=1}^{n}1_{k_{m}>k}e^{-pk_{m}}=\left( \sum_{l>k}e^{-pl}\right)
^{n}=\left( \frac{e^{-p\left( k+1\right) }}{1-e^{-p}}\right) ^{n} 
\end{equation*}
is the normalizing constant.

Defining $G_{n,p}:=-\log Z_{n,p}$, we have $\partial _{p}G_{n,p}=\Bbb{E}%
_{N,p}\left( \sum \widetilde{N}_{m,n}1_{\widetilde{N}_{m,n}>k}\right) $ and
one must choose $p$ in such a way that $\partial _{p}G_{n,p}=N$, leading to $%
N=n\left( k+1+\frac{e^{-p}}{1-e^{-p}}\right) $or $\frac{1}{\rho }=\frac{%
e^{-p}}{1-e^{-p}}+k+1$, so $p=-\log \left( \frac{1-\rho \left( k+1\right) }{%
1-\rho k}\right) .$ The latter equation relating $p,\rho $ and $k$ is an
equation of state. Due to the equivalence of ensembles principle, see \cite
{Hui2} for similar arguments, we have: $Z_{n,N}=e^{pN}Z_{n,p}O\left(
N^{-1/2}\right) ,$ leading to: $-\frac{1}{n}\log Z_{n,N}\sim -\frac{1}{n}%
\log Z_{n,p}-\frac{p}{\rho }$. Proceeding in this way, we finally get 
\begin{equation}
-\frac{1}{n}\log \mathbf{P}\left( N_{1:n}>k\right) \rightarrow  \label{b2}
\end{equation}
\begin{equation*}
F_{hr}\left( p,\rho \right) =-\frac{1}{\rho }\left( \rho \log \rho +\left(
1-\rho \right) \log \left( 1-\rho \right) \right) -\frac{p}{\rho }-\log
\left( \frac{e^{-p\left( k+1\right) }}{1-e^{-p}}\right) , 
\end{equation*}
with $\rho \in \left( 0,1/\left( k+1\right) \right) $. Here, thermodynamical
``pressure'' $p>0$ and density $\rho $ are related through the ``state
equation'' $\partial _{p}F_{hr}\left( p,\rho \right) =0$ which can
consistently be checked to be 
\begin{equation}
\frac{1}{\rho }=k+1+\frac{e^{-p}}{1-e^{-p}},  \label{b3}
\end{equation}
leading to $p=-\log \left( \frac{1-\rho \left( k+1\right) }{1-\rho k}\right)
>0$ (which is well-defined and positive because $\rho <1/\left( k+1\right) $%
). Thus $F_{hr}$ is an explicit entropy-like positive function of $\rho $
and $k$, namely 
\begin{equation}
F_{hr}\left( \rho \right) =  \label{b4}
\end{equation}
\begin{equation*}
-\frac{1}{\rho }\left( \left( 1-\rho \right) \log \left( 1-\rho \right)
-\left( 1-\rho \left( k+1\right) \right) \log \left( 1-\rho \left(
k+1\right) \right) +\left( 1-\rho k\right) \log \left( 1-\rho k\right)
\right) 
\end{equation*}

In the thermodynamical limit, hard-rods configurations are exceptional and
the hard-rods large deviation rate function $F_{hr}$ is an explicit function
of $\rho $ and $k.$ We conclude that with probability tending to $1$, $%
N_{1:n}=1$: In the partitioning approach of the fortune $N$ amongst $n$
recipients, the share of the poorer is the smallest possible$.$

As $\rho \uparrow 1/\left( k+1\right) $, pressure tends to $\infty $ and $%
F_{hr}\left( \rho \right) \rightarrow \left( k+1\right) \log \left(
k+1\right) -k\log \left( k\right) >0.$ As $\rho \downarrow 0$, pressure
tends to $0$ and $F_{hr}\left( \rho \right) \rightarrow 0.$

\subsection{Covering configurations}

Covering configurations are those for which we have $\left( N_{n:n}\leq
k\right) .$ In the partitioning approach of the fortune $N$ amongst $n$
recipients, this event is realized when the share of the richest is bounded
above by $k$ (a rare event). Assume $n,N\rightarrow \infty $ with $%
n/N\rightarrow \rho \in \left( 1/k,1\right) $ where $k>1$ is fixed. One also
expects that the probability of covering configurations by arcs of length $k$
tends to zero exponentially fast. Working now with 
\begin{equation*}
Z_{n,p}=\sum_{k_{1},..,k_{n}\geq 1}\prod_{m=1}^{n}1_{k_{m}\leq
k}e^{-pk_{m}}=\left( e^{-p}\frac{1-e^{-pk}}{1-e^{-p}}\right) ^{n}
\end{equation*}
and proceeding as for the hard rods case, we easily get 
\begin{equation*}
-\frac{1}{n}\log \mathbf{P}\left( N_{n:n}\leq k\right) \rightarrow
F_{c}\left( p,\rho \right) 
\end{equation*}
where the covering large deviation rate function is 
\begin{equation}
F_{c}\left( p,\rho \right) =-\frac{1}{\rho }\left( \rho \log \rho +\left(
1-\rho \right) \log \left( 1-\rho \right) \right) -\frac{p}{\rho }-\log
\left( \frac{e^{-p}\left( 1-e^{-pk}\right) }{1-e^{-p}}\right) .  \label{b5}
\end{equation}
Here, thermodynamical pressure $p$ and density $\rho \in \left( 1/k,1\right) 
$ are related through the covering state equation $\partial _{p}F_{c}\left(
p,\rho \right) =0$, namely 
\begin{equation}
\frac{1}{\rho }=1+\frac{e^{-p}}{1-e^{-p}}-\frac{ke^{-pk}}{1-e^{-pk}}.
\label{b6}
\end{equation}
For all finite arc-length $k,$ $k-$covering configurations are also
exceptional. The $k-$covering large deviation rate function $F_{c}$ is in
general an implicit function of $\rho $ and $k$, $\rho \in \left(
1/k,1\right) .$ When $\rho \downarrow 1/k$, pressure tends to $-\infty $ and 
$F_{c}\left( \rho \right) \rightarrow k\log k-\left( k-1\right) \log \left(
k-1\right) >0.$ As $\rho \uparrow 1$, pressure tends to $\infty $ and $%
F_{c}\left( \rho \right) \rightarrow 0.$ By continuity, there is a value of $%
\rho _{0}$ inside the definition domain of $\rho $ where $p=0.$ We have $%
F_{c}\left( p,\rho _{0}\right) =-\frac{1}{\rho _{0}}\left( \rho _{0}\log
\rho _{0}+\left( 1-\rho _{0}\right) \log \left( 1-\rho _{0}\right) \right)
-\log k.$ In the partitioning approach of the fortune $N$ amongst $n$
recipients, the share of the richest is bounded above with probability
tending to $0$ exponentially fast.\newline

\textbf{Remark:} When $k=2,$ the covering equation of state can be solved
explicitly because it boils down to a second degree equation in $e^{-p}.$
One finds $p=-\log \left( \frac{1-\rho }{2\rho -1}\right) $. Plugging in
this expression of $p$ in $F_{c}\left( p,\rho \right) $ with $k=2$ gives 
\begin{equation*}
F_{c}=-\frac{1}{\rho }\left( 2\rho \log \rho -\left( 2\rho -1\right) \log
\left( 2\rho -1\right) \right) , 
\end{equation*}
an explicit function of $\rho \in \left( 1/2,1\right) .$ Note that $%
p\uparrow \infty $ as $\rho \uparrow 1,$ $p\downarrow -\infty $ as $\rho
\downarrow 1/2$ and $p=0$ when $\rho =2/3.$ We have $F_{c}\left(
p,2/3\right) =\frac{3}{2}\log 3-2\log 2.$

\subsection{Parking configurations}

Parking configurations are those for which we have $\left(
N_{1:n}>k,N_{n:n}\leq 2k\right) .$ In the partitioning approach of the
fortune $N$ amongst $n$ recipients, this event is realized if the share of
the richest is bounded above by twice the share of the poorest. Assume $%
n,N\rightarrow \infty $ with $n/N\rightarrow \rho \in \left( 1/\left(
2k\right) ,1/\left( k+1\right) \right) $. One expects that the probability
of $k-$parking configurations tends to zero exponentially fast. Working now
with 
\begin{equation*}
Z_{n,p}=\sum_{k_{1},..,k_{n}\geq 1}\prod_{m=1}^{n}1_{k<k_{m}\leq
2k}e^{-pk_{m}}=\left( e^{-p\left( k+1\right) }\frac{1-e^{-pk}}{1-e^{-p}}%
\right) ^{n}
\end{equation*}
and proceeding as for the hard rods case, we easily get 
\begin{equation*}
-\frac{1}{n}\log \mathbf{P}\left( N_{1:n}>k,N_{n:n}\leq 2k\right)
\rightarrow F_{\pi }\left( p,\rho \right) 
\end{equation*}
where the parking large deviation rate function is 
\begin{equation}
F_{\pi }\left( p,\rho \right) =-\frac{1}{\rho }\left( \rho \log \rho +\left(
1-\rho \right) \log \left( 1-\rho \right) \right) -\frac{p}{\rho }-\log
\left( \frac{e^{-p\left( k+1\right) }\left( 1-e^{-pk}\right) }{1-e^{-p}}%
\right) .  \label{b7}
\end{equation}
Here, thermodynamical pressure $p$ and density $\rho \in \left( 1/k,1\right) 
$ are related through the parking equation of state $\partial _{p}F_{\pi
}\left( p,\rho \right) =0$, namely 
\begin{equation}
\frac{1}{\rho }=k+1+\frac{e^{-p}}{1-e^{-p}}-\frac{ke^{-pk}}{1-e^{-pk}}.
\label{b8}
\end{equation}
The parking configurations large deviation rate function $F_{\pi }$ is an
implicit function of $\rho $ and $k$ with $\rho \in \left( 1/\left(
2k\right) ,1/\left( k+1\right) \right) .$ The latter formula can be extended
to the border case $k=1$. Indeed, when $k=1$, $\rho =1/2,$ pressure tends to 
$\infty $ and $F_{\pi }\left( p,\rho \right) =2\log 2.$ From Stirling
formula, this is in agreement with the fact $\mathbf{P}\left( N_{1:n}>1,%
\text{ }N_{n:n}\leq 2\right) =\binom{N-1}{n-1}^{-1}\neq 0$ only if $N=2n$,
which is the probability of the regular configuration where the $n$ sampled
points are all exactly equally-spaced by two arc length units.\newline

\textbf{Remark:} When $k=2,$ the parking equation of state can be solved
explicitly to give $p=-\log \left( \frac{1-3\rho }{4\rho -1}\right) $.
Plugging in this expression of $p$ into $F_{\pi }\left( p,\rho \right) $
with $k=2$ gives $F_{\pi }$ as an explicit function of $\rho \in \left(
1/4,1/3\right) .$ Note that $p\uparrow \infty $ as $\rho \uparrow 1/3,$ $%
p\downarrow -\infty $ as $\rho \downarrow 1/4$ and $p=0$ when $\rho =2/7.$

\section{The grand canonical partition of $N$}

Suppose $N$ indistinguishable balls are assigned at random into $N$
indistinguishable boxes. Let $N_{n,N}\geq 0$ be the number of balls in box
number $n.$ This leads to a random partition of $N$ now into $N$ id summands
which are $\geq 0:$%
\begin{equation}
N=\sum_{n=1}^{N}N_{n,N}.  \label{c1}
\end{equation}
We have 
\begin{equation}
\mathbf{P}\left( N_{1,N}=k_{1},..,N_{N,N}=k_{N}\right) =\frac{1}{\binom{2N-1%
}{N}},  \label{c2}
\end{equation}
which is a Bose-Einstein distribution on the full $N-$simplex: 
\begin{equation*}
\left\{ k_{n}\geq 0\text{ satisfying }\sum_{n=1}^{N}k_{n}=N\right\} . 
\end{equation*}
Summing over all the $k_{n}$ but one, the marginal distribution of $N_{1,N}$
is easily seen to be 
\begin{equation}
\mathbf{P}\left( N_{1,N}=k\right) =\frac{\binom{2N-k-2}{N-k}}{\binom{2N-1}{N}%
},\text{ }k=0,..,N.  \label{c3}
\end{equation}
Let $P_{N}=\sum_{n=1}^{N}1\left( N_{n,N}>0\right) $ count the number of
summands which are strictly positive (the number of non-empty boxes). With $%
k_{m}\geq 1$ satisfying $\sum_{m=1}^{n}k_{m}=N$, we obtain 
\begin{equation}
\mathbf{P}\left( N_{1,N}=k_{1},..,N_{n,N}=k_{n};P_{N}=n\right) =\frac{\binom{%
N}{n}}{\binom{2N-1}{N}},  \label{c4}
\end{equation}
which is independent of the filled box occupancies $\left(
k_{1},..,k_{n}\right) $ (the probability being uniform).

As there are $\binom{N-1}{n-1}$ sequences $k_{m}\geq 1$, $m=1,..,n$
satisfying $\sum_{m=1}^{n}k_{m}=N$, summing over the $k_{m}\geq 1,$ we get
the hypergeometric distribution for $P_{N}:$%
\begin{equation}
\mathbf{P}\left( P_{N}=n\right) =\frac{\binom{N}{n}\binom{N-1}{n-1}}{\binom{%
2N-1}{N}},\text{ }n=1,..,N.  \label{c5}
\end{equation}
This distribution occurs in the following urn model: Draw $N$ balls without
replacement from an urn containing $2N-1$ balls in total, $N$ of which are
white, $N-1$ are black. The law of $P_{N}$ describes the distribution of the
number of white balls drawn from the urn. Its mean is $N^{2}/\left(
2N-1\right) \sim N/2$ and its variance is $\left( N^{2}\left( N-1\right)
\right) /\left( 2\left( 2N-1\right) ^{2}\right) \sim N/8.$

As a result, 
\begin{equation}
\mathbf{P}\left( N_{1,N}=k_{1},..,N_{n,N}=k_{n}\mid P_{N}=n\right) =\frac{1}{%
\binom{N-1}{n-1}}1\left( \left| \mathbf{k}\right| =N\right)  \label{c6}
\end{equation}
which is the spacings conditional Bose-Einstein model with $\mathbf{k}\geq 
\mathbf{1}$ described in (\ref{a3}). The balls in boxes model just defined
is therefore an extension of the conditional Bose-Einstein model allowing
the number of sampled points to be unknown and random.\newline

\textbf{Repetitions (grand canonical).} It is likely that some boxes contain
the same number of particles. To take these multiplicities into account, let 
$A_{i,N}$, $i\in \left\{ 0,..,N\right\} $ count the number of boxes with
exactly $i$ balls, that is 
\begin{equation}
A_{i,N}=^{\#}\left\{ n\in \left\{ 1,..,N\right\} :N_{n,N}=i\right\}
=\sum_{n=1}^{N}1\left( N_{n,N}=i\right) .  \label{c7}
\end{equation}
Then $\sum_{i=0}^{N}A_{i,N}=N$ where $\sum_{i=1}^{N}A_{i,N}=P_{N}$ is the
number of filled boxes and $A_{0,N}=N-P_{N}$ the number of empty ones. The
joint probability of the $A_{i,N}$ is given by the Ewens formula (see \cite
{Ewens} and \cite{Hui3}) 
\begin{equation}
\mathbf{P}\left( A_{0,N}=a_{0},A_{1,N}=a_{1},..,A_{N,N}=a_{N}\right) =\frac{1%
}{\binom{2N-1}{N}}\frac{N!}{\prod_{i=0}^{N}a_{i}!},  \label{c8}
\end{equation}
on the set $\sum_{i=0}^{N}a_{i}=\sum_{i=1}^{N}ia_{i}=N.$\newline

Let us now investigate the marginal law of the $A_{i,N}.$ Firstly, the law
of $A_{0,N}=N-P_{N}$ clearly is 
\begin{equation}
\mathbf{P}\left( A_{0,N}=a_{0}\right) =\frac{\binom{N}{a_{0}}\binom{N-1}{%
a_{0}}}{\binom{2N-1}{N}},\text{ }a_{0}=0,..,N-1,  \label{c9}
\end{equation}
with $\mathbf{E}\left( A_{0,N}\right) \sim N/2$. Secondly, recalling $%
A_{i,N}=\sum_{n=1}^{N}1\left( N_{n,N}=i\right) ,$ with $\left( N\right)
_{l}:=N\left( N-1\right) ..\left( N-l+1\right) $, using the exchangeability
of $\left( N_{1,N},..,N_{N,N}\right) $, the probability generating function
of $A_{i,N}$ ($i\neq 0$) reads 
\begin{equation*}
\mathbf{E}\left( z^{A_{i,N}}\right) =1+\sum_{l\geq 1}\frac{\left( z-1\right)
^{l}}{l!}\left( N\right) _{l}\mathbf{P}\left( N_{1,N}=i,..,N_{l,N}=i\right)
. 
\end{equation*}
Using $\mathbf{P}\left( N_{1,N}=k_{1},..,N_{l,N}=k_{l}\right) =\binom{%
2N-l-\sum_{1}^{l}k_{m}-1}{N-l-1}/\binom{2N-1}{N-1}$, we get the falling
factorial moments of $A_{i,N}$ as 
\begin{equation}
m_{l,i}\left( N\right) :=\mathbf{E}\left[ \left( A_{i,N}\right) _{l}\right]
=\left( N\right) _{l}\binom{2N-l-li-1}{N-l-1}/\binom{2N-1}{N-1},  \label{c10}
\end{equation}
where $l\in \left\{ 0,..,l\left( i\right) =\left( N-1\right) \wedge \left[
N/i\right] \right\} .$ The marginal distribution of $A_{i,N}$ is thus 
\begin{equation}
\mathbf{P}\left( A_{i,N}=a_{i}\right) =\sum_{l=a_{i}}^{l\left( i\right) }%
\frac{(-1)^{l-a_{i}}}{l!}\binom{l}{a_{i}}m_{l,i}\left( N\right) ,\text{ }%
a_{i}\in \left\{ 0,..,l\left( i\right) \right\} .  \label{c11}
\end{equation}
If $l=1$, $\mathbf{E}\left( A_{i,N}\right) =N\binom{2N-i-2}{N-2}/\binom{2N-1%
}{N-1}.$ The variance of $A_{i,N}$ is $\sigma ^{2}\left( A_{i,N}\right)
=m_{2,i}\left( N\right) +m_{1,i}\left( N\right) -m_{1,i}\left( N\right)
^{2}. $\newline

In particular, we find that $\mathbf{E}\left( A_{1,N}\right) =N\left(
N+1\right) /\left( 2\left( 2N+1\right) \right) \sim N/4$ is the mean number
of singleton boxes in the grand canonical model: when $N$ is large, about
one fourth out of the $N$ boxes is filled by singletons (recall that one
half of $N$ is filled by no ball). The variance of $A_{1,N}$ is $\sigma
^{2}\left( A_{1,N}\right) \sim N/4$ so we expect that $A_{1,N},$ properly
normalized, converges to a normal distribution$.$ Next, we can check that $%
\mathbf{E}\left( A_{2,N}\right) \sim N/8$ and further that $\mathbf{E}\left(
A_{i,N}\right) \sim N/2^{i+1},$ showing a geometric decay in $i$ of $\mathbf{%
E}\left( A_{i,N}\right) .$

Finally, note that the probability that $A_{i,N}$ takes its maximal possible
value $l\left( i\right) $ is 
\begin{equation*}
\mathbf{P}\left( A_{i,N}=l\left( i\right) \right) =m_{l\left( i\right)
,i}\left( N\right) /l\left( i\right) !=\binom{N}{l\left( i\right) }/\binom{%
2N-1}{N-1}. 
\end{equation*}
For example $\mathbf{P}\left( A_{1,N}=N-1\right) =N/\binom{2N-1}{N-1}$ is
the (exponentially small) probability that all $N$ boxes are filled by
singletons.\newline

\textbf{Multiplicities and conditioning.} Let us now investigate the same
problem while conditioning on $P_{N}=n.$

Firstly, note that $\sum_{i=1}^{N}iA_{i,N}\left( i\right) =N$ is the total
number of balls. Using the multinomial formula, with $\sum_{i=1}^{N}ia_{i}=N$
and $\sum_{i=1}^{N}a_{i}=n$, we thus get 
\begin{equation}
\mathbf{P}\left( A_{1,N}=a_{1},..,A_{N,N}=a_{N},P_{N}=n\right) =\frac{\binom{%
N}{n}}{\binom{2N-1}{N}}\frac{n!}{\prod_{i=1}^{N}a_{i}!}  \label{c12}
\end{equation}
and 
\begin{equation}
\mathbf{P}\left( A_{1,N}=a_{1},..,A_{N,N}=a_{N}\mid P_{N}=n\right) =\frac{n!%
}{\binom{N-1}{n-1}}\frac{1}{\prod_{i=1}^{N}a_{i}!}.  \label{c13}
\end{equation}
The latter formulae give the joint (Ewens-like) distributions of the
repetition vector count.\newline

Let us investigate the marginal distribution of the $A_{i,N}$ conditional
given $P_{N}=n$. Firstly, the law of $A_{0,N}=N-P_{N}$ is $\mathbf{P}\left(
A_{0,N}=a_{0}\mid P_{N}=n\right) =\delta _{a_{0}-\left( N-n\right) }.$%
\newline

Secondly, recalling $A_{i,N}=\sum_{n=1}^{N}1\left( N_{n,N}=i\right) ,$ with $%
\left( n\right) _{l}=n\left( n-1\right) ..\left( n-l+1\right) $ (and $\left(
n\right) _{0}:=1$)$,$ using the exchangeability of $\left(
N_{1,N},..,N_{N,N}\right) $, the conditional probability generating function
of $A_{i,N}$ reads 
\begin{equation*}
\mathbf{E}\left( z^{A_{i,N}}\mid P_{N}=n\right) =1+\sum_{l\geq 1}\frac{%
\left( z-1\right) ^{l}}{l!}\left( n\right) _{l}\mathbf{P}\left(
N_{1,N}=i,..,N_{l,N}=i\mid P_{N}=n\right) . 
\end{equation*}
Using $\mathbf{P}\left( N_{1,N}=k_{1},..,N_{l,N}=k_{l}\mid P_{N}=n\right) =%
\binom{N-\sum_{1}^{l}k_{m}-1}{n-l-1}/\binom{N-1}{n-1}$, we get the
conditional falling factorial moments of $A_{i,N}$ as 
\begin{equation}
m_{l,i}\left( n,N\right) :=\mathbf{E}\left[ \left( A_{i,N}\right) _{l}\mid
P_{N}=n\right] =\left( n\right) _{l}\binom{N-li-1}{n-l-1}/\binom{N-1}{n-1},
\label{c14}
\end{equation}
where $l\in \left\{ 0,..,l\left( i\right) =\left( n-1\right) \wedge \left[
\left( N-1\right) /i\right] \right\} .$ The conditional marginal
distribution of $A_{i,N}$ is thus 
\begin{equation}
\mathbf{P}\left( A_{i,N}=a_{i}\mid P_{N}=n\right) =\sum_{l=a_{i}}^{l\left(
i\right) }\frac{(-1)^{l-a_{i}}}{l!}\binom{l}{a_{i}}m_{l,i}\left( n,N\right) .
\label{c15}
\end{equation}

If $l=1$, $\mathbf{E}\left( A_{i,N}\mid P_{N}=n\right) =n\binom{N-i-1}{n-2}/%
\binom{N-1}{n-1}.$ In particular, $\mathbf{E}\left( A_{1,N}\mid
P_{N}=n\right) =n\left( n-1\right) /\left( N-1\right) $ is the mean number
of singleton boxes. In the thermodynamical limit $n,N\rightarrow \infty $, $%
n/N\rightarrow \rho $, $\mathbf{E}\left( A_{1,N}\mid P_{N}=n\right) \sim
\rho n$ and a fraction $\rho $ of the $n$ filled boxes is filled with
singletons. For the variance, we have $\sigma ^{2}\left( A_{1,N}\mid
P_{N}=n\right) \sim \rho n.$ We can also check that a fraction $\rho \left(
1-\rho \right) $ of the $n$ filled boxes is filled with doubletons: $\mathbf{%
E}\left( A_{2,N}\mid P_{N}=n\right) \sim \rho \left( 1-\rho \right) n$ and
more generally that $\mathbf{E}\left( A_{i,N}\mid P_{N}=n\right) \sim \rho
\left( 1-\rho \right) ^{i}n.$

Finally, note that the probability that $A_{i,N}$ reaches its maximal
possible value $l\left( i\right) $ is 
\begin{equation*}
\mathbf{P}\left( A_{i,N}=l\left( i\right) \mid P_{N}=n\right) =m_{l\left(
i\right) ,i}\left( n,N\right) /l\left( i\right) !=\binom{n}{l\left( i\right) 
}/\binom{N-1}{n-1}. 
\end{equation*}
For example $\mathbf{P}\left( A_{1,N}=n-1\mid P_{N}=n\right) =n/\binom{N-1}{%
n-1} $ is the probability that $n-1$ boxes are filled by singletons and one
box by $N-n+1$ balls, which is obvious.

\section{Random Graph Connectivity}

The latter model may be viewed as a clockwise $k-$nearest neighbor graph
with $N$ vertices and $kn$ edges. Consider as before $N$ equally spaced
points (vertices) on the $N-$circle so with arc length $1$ between
consecutive points. Draw at random $n\in \left\{ 2,..,N-1\right\} $ points
without replacement at the integer vertices of this circle. Assume $N\leq 2n$
and draw an edge at random from each of the $n$ sampled points, removing
each sampled point once it has been paired. At the end of this process, we
get a random graph with $N$ vertices and $n$ (out-degree $1$) edges whose
endpoints are no more neighbors, being now chosen at random on $\left\{
1,..,N\right\} $. We wish to estimate the covering probability for this new
model in the spirit of Erd\H{o}s-R\'{e}nyi random graphs.\newline

Let $B_{m}$, $m=1,..,n$ be a sequence of independent (but not id) Bernoulli
rvs with success probabilities $p_{m}=\frac{N-n}{N-\left( m-1\right) },$ $%
m=1,..,n.$ With $\left[ z^{k}\right] \phi \left( z\right) $ the $z^{k}-$%
coefficient of $\phi \left( z\right) ,$ the $N-$covering probability is 
\begin{equation}
\mathbf{P}_{n,N}\left( \text{cover}\right) =\mathbf{P}\left( N-n\leq
\sum_{m=1}^{n}B_{m}\leq n\right) =\sum_{k=N-n}^{n}\left[ z^{k}\right] 
\mathbf{E}\left( z^{\sum_{m=1}^{n}B_{m}}\right) ,  \label{d1}
\end{equation}
which is just the probability to hit all points of the un-sampled set $%
\left\{ n+1,..,N\right\} $ at least once in a uniform pairing without
replacement of the $n-$sample$.$ This covering probability is the
probability of connectedness of the random graph with $N$ vertices and $n$
out-degree $1$ edges. It is of course zero if $N>2n.$ Let $\overline{p}_{n}=%
\frac{1}{n}\sum_{m=1}^{n}p_{m}$ be the sample mean of the Bernoulli rvs. The
covering probability can be bounded by 
\begin{equation}
\mathbf{P}_{n,N}\left( \text{cover}\right) \leq \sum_{k=N-n}^{n}\binom{n}{k}%
\left( 1-\overline{p}_{n}\right) ^{k}\overline{p}_{n}^{n-k}.  \label{d2}
\end{equation}
Assume $n,N\rightarrow \infty $ while $n/N\rightarrow \rho $ so with $\rho
\in \left( 1/2,1\right) .$ Then 
\begin{equation*}
\overline{p}_{n}\rightarrow -\frac{1-\rho }{\rho }\log \left( 1-\rho \right)
=:\mu \left( \rho \right) . 
\end{equation*}
Clearly $\sigma ^{2}\left( B_{m}\right) <\infty $ and $\sum_{m=1}^{n}m^{-2}%
\sigma ^{2}\left( B_{m}\right) $ has a finite limit. By Kolmogorov Strong
Law of Large Numbers $\frac{1}{n}\sum_{m=1}^{n}B_{m}\overset{a.s.}{%
\rightarrow }\mu \left( \rho \right) $ and so $\mathbf{P}_{N}\left( \text{%
cover}\right) \rightarrow 1$ if $\rho \geq \rho _{c}:=1-e^{-1}$ because in
this case the probability to estimate is 
\begin{equation*}
\mathbf{P}\left( \frac{1}{n}\sum_{m=1}^{n}B_{m}\geq \frac{1-\rho }{\rho }%
\right) , 
\end{equation*}
with $\mu \left( \rho \right) \in \left[ \frac{1-\rho }{\rho },1\right] $.

Whereas, when $\rho <1-e^{-1},$ the bound for the covering probability can
be estimated by 
\begin{equation*}
\mathbf{P}_{n,N}\left( \text{cover}\right) \leq N\int_{1-\rho }^{\rho }%
\binom{N\rho }{Nx}\left( 1-\mu \left( \rho \right) \right) ^{Nx}\mu \left(
\rho \right) ^{N\left( \rho -x\right) }dx 
\end{equation*}
\begin{equation*}
\sim CN\int_{1-\rho }^{\rho }e^{NH_{\rho }\left( x\right) }dx, 
\end{equation*}
where $H_{\rho }\left( x\right) =\rho \log \rho -x\log x-\left( \rho
-x\right) \log \left( \rho -x\right) +x\log \left( 1-\mu \left( \rho \right)
\right) +\left( \rho -x\right) \log \mu \left( \rho \right) .$ The function $%
x\rightarrow H_{\rho }\left( x\right) $ is concave and attains its maximum
at $x=\rho \left( 1-\mu \left( \rho \right) \right) <1-\rho ,$ which is
outside the integration interval $\left[ 1-\rho ,\rho \right] $. By the
saddle point method 
\begin{equation}
\underset{n,N\rightarrow \infty ,\text{ }n/N\rightarrow \rho }{\lim \inf }-%
\frac{1}{n}\log \mathbf{P}_{n,N}\left( \text{cover}\right) \underset{}{=}%
F_{G}\left( \rho \right) :=-\frac{1}{\rho }H_{\rho }\left( 1-\rho \right) >0.
\label{d3}
\end{equation}
So only in the low-density range $\frac{1}{2}<\rho <1-e^{-1}$ is the graph's
connectedness probability exponentially small. Note that the graph large
deviation rate function $F_{G}$ is maximal (minimal) at $\rho =1/2$ ($\rho
_{c}=1-e^{-1}$), with $F_{G}\left( \rho _{c}\right) =-\frac{3-e}{e-1}\log
\left( e-2\right) >0.$\newline

We conclude that in the random graph approach to the covering problem, in
sharp contrast to the $k-$nearest neighbor graph, there exists a critical
density $\rho _{c}=1-e^{-1}$ above which covering occurs with probability
one. These results illustrate to what extent, when connections are not
restricted to neighbors, the chance of connectedness is increased. This
question was also raised in (\cite{Can}, p.$18$) in relation with
Small-World graphs.

\end{document}